\documentclass[a4paper,fleqn,usenatbib]{mnras}

\usepackage{newtxtext,newtxmath}

\usepackage[T1]{fontenc}
\usepackage{ae,aecompl}

\usepackage{graphicx}	
\usepackage{amsmath}	
\usepackage{amssymb}	
\usepackage{mathtools}
\usepackage{setspace}
\usepackage{tabularx}
\usepackage{tabulary}

\DeclarePairedDelimiter{\ceil}{\lceil}{\rceil}
\DeclarePairedDelimiter{\floor}{\lfloor}{\rfloor}
\newcommand{\changes}[1]{{#1}}

\title[Modelling the structure of star clusters]{Modelling the structure of star clusters with fractional Brownian motion}

\author[O. Lomax, M. L. Bates \& A. P. Whitworth]{O. Lomax\thanks{E-mail: oliver.lomax@astro.cf.ac.uk}, M. L. Bates, \& A. P. Whitworth\\
School of Physics and Astronomy, Cardiff University, Cardiff CF24 3AA, UK}

\date{Accepted XXX. Received YYY; in original form ZZZ}

\pubyear{2018}

\begin{document}
\label{firstpage}
\pagerange{\pageref{firstpage}--\pageref{lastpage}}
\maketitle

\begin{abstract}


\noindent{}The degree of fractal substructure in molecular clouds can be quantified by comparing them with Fractional Brownian Motion (FBM) surfaces or volumes. These fields are self-similar over all length scales and characterised by a drift exponent $H$, which describes the structural roughness. Given that the structure of molecular clouds and the initial structure of star clusters are almost certainly linked, it would be advantageous to also apply this analysis to clusters. Currently, the structure of star clusters is often quantified by applying $\mathcal{Q}$ analysis. $\mathcal{Q}$ values from observed targets are interpreted by comparing them with those from artificial clusters. These are typically generated using a Box-Fractal (BF) or Radial Density Profile (RDP) model. We present a single cluster model, based on FBM, as an alternative to these models. Here, the structure is parameterised by $H$, and the standard deviation of the log-surface/volume density $\sigma$. The FBM model is able to reproduce both centrally concentrated and substructured clusters, and is able to provide a much better match to observations than the BF model. We show that $\mathcal{Q}$ analysis is unable to estimate FBM parameters. Therefore, we develop and train a machine learning algorithm which can estimate values of $H$ and $\sigma$, with uncertainties. This provides us with a powerful method for quantifying the structure of star clusters in terms which relate to the structure of molecular clouds. We use the algorithm to estimate the $H$ and $\sigma$ for several young star clusters, some of which have no measurable BF or RDP analogue.



\end{abstract}




\section{Introduction}


\changes{Recent spaceborne instruments have revealed much of the detailed multiscale structure of our own galaxy. The \textsc{Herschel} submillimetre observatory \citep{spire,pacs} has mapped out many of the gas and dust structures in the InterStellar Medium (ISM) \citep[e.g.][]{higal}. Similarly, the \textsc{Gaia} observatory \citep{gaia,gaia_DR2} continues to reveal the spatial and velocity distribution of the stars which accompany this gas and dust. Nevertheless, understanding the link between the structures in the ISM and star clusters remains an ongoing challenge}. We are confident that the earliest stages of stellar evolution occur within dense, substructured (i.e. clumpy or filamentary) molecular clouds within the ISM \citep[e.g.][]{MAN98,AMB10,SGKF16,P18}. However, the extent to which star clusters retain the structural signatures of their parent molecular clouds is uncertain. Some studies highlight similarities between the distribution of stars and that of the molecular clouds which spawn them \citep[e.g.][]{EF96,GHK14}. However, numerical studies suggest that gas and stars decouple quickly during the star formation process, erasing structural similarities \citep[e.g.][]{BB05,PD15}. \changes{To make headway in this complex field, we require tools which can fulfil two roles. First, we need statistics which can quantify the structure of clouds and clusters, ideally in the same terms. Second, in order to simulate these structures, we need initial conditions which statistically match observations.}

\citet{SBH98} note that molecular clouds can be compared with surface-density fields generated by Fractional Brownian Motion (FBM). These are random fractal structures, with well defined fractal dimension $D$, which can be analysed using perimeter-area or $\Delta$-variance techniques \citep[e.g.][]{FPW91,SBH98,WBM00,ESS14}. Other studies measure the surface-density Probability Density Functions (PDFs) of molecular clouds \citep[e.g.][]{FK12,SAK13}. These can provide a measure of a cloud's surface-density dynamic range, which is not necessarily related to its fractal structure. Indeed, a property of fractal distributions is that the density can be rescaled by any one-to-one transform without altering $D$ \citep{PS88}.

Techniques also exist which estimate the fractal properties of star clusters. \citet[][hereafter CW04]{CW04} were the first to use minimum spanning trees to estimate $D$ for clusters. The application of this method has since become widespread in the field star formation \citep[e.g.][]{SK06,C09a,CW09,LWC11,PWG14,P18}. However, this analysis assumes that substructured clusters can be described by a Box-Fractal (BF) \citep{GW04} or a Radial Density Profile (RDP) model. The BF model is parameterised by $D$ only. Here, altering $D$ also changes the surface-density dynamic range; the two properties cannot be varied independently. A more recent study by \citet[][hereafter JWL17]{JWL17} expands the BF model to include variable surface-density scaling. This model provides a better likeness to observed clusters, at the cost of two additional parameters.

In this paper, we present a method of generating model FBM star clusters. \changes{This provides a parameterisation of cluster structure which matches that of clouds}. We demonstrate that BF clusters do not always match observations, and therefore should not be used to infer quantitative results. We show that FBM clusters overcome this problem and we use machine learning to estimate the structural parameters of test clusters and observations. In Section \ref{sec:model_star_clusters} of the paper we define different star cluster models. In Section \ref{sec:parameter_estimators} we review parameter estimators and apply them to observations. In Section \ref{sec:discussion} we compare and discuss the results of the estimators. Finally, we summarise our conclusions in Section \ref{sec:conclusions}.


\section{Model star clusters}
\label{sec:model_star_clusters}

\begin{figure*}
\centering
\includegraphics[width=0.49\textwidth]{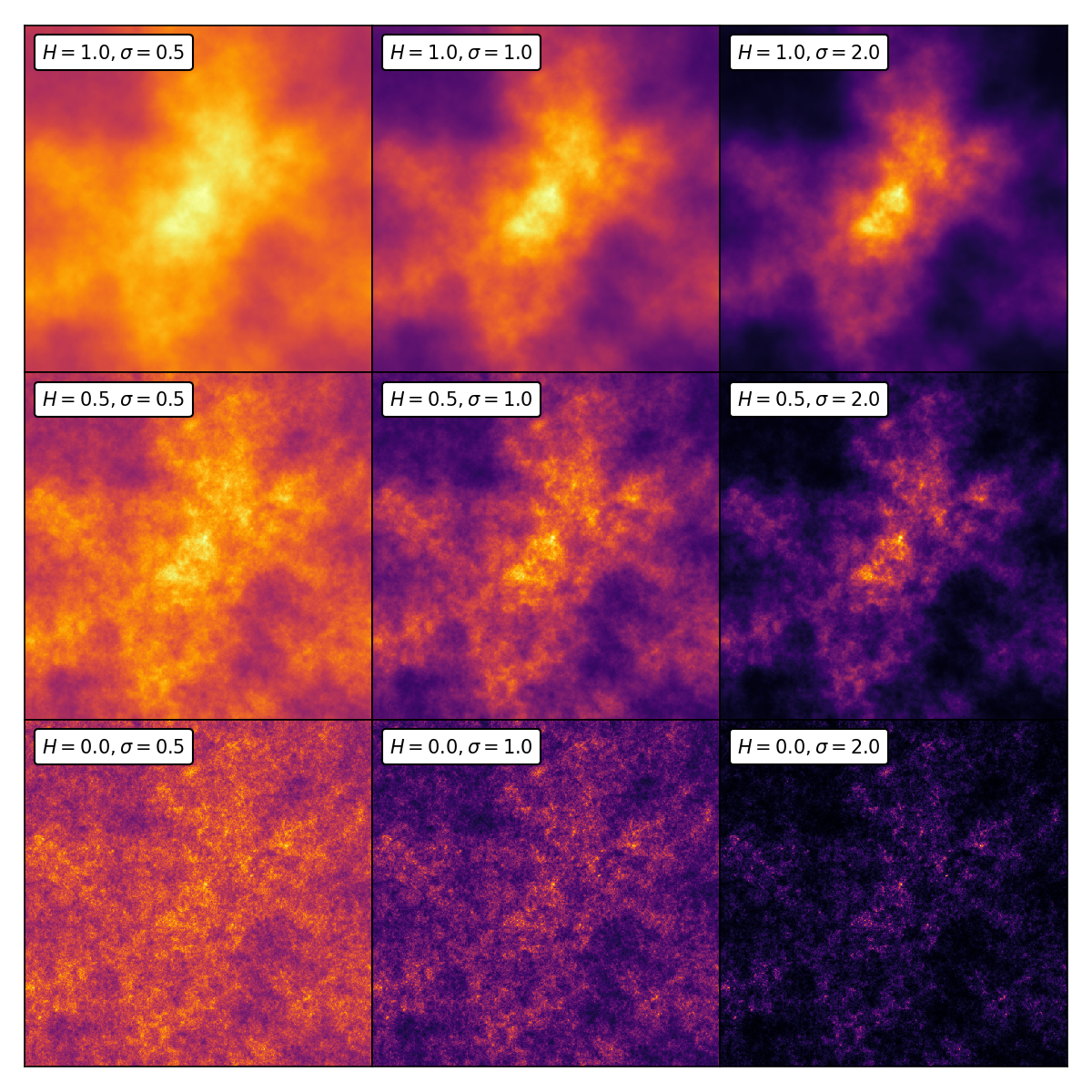}
\includegraphics[width=0.49\textwidth]{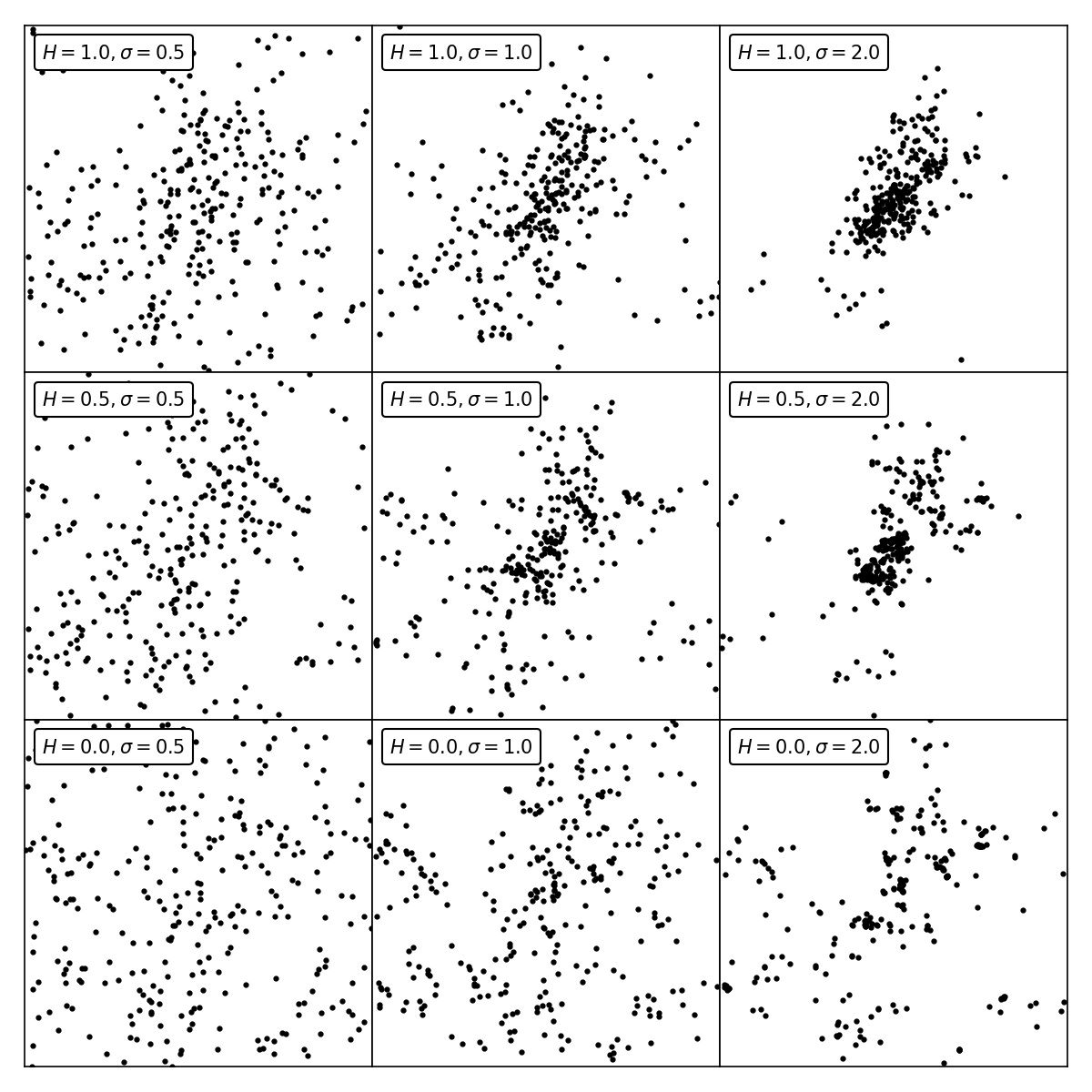}
\caption{Left: nine $E_2$ FBM fields generated using the same random seed. From top-to-bottom, the rows show fields with $H=1.0,\,0.5,\text{ and }0.0$. From left-to-right, the columns show fields with $\sigma=1.0\,,2.0\text{ and }3.0$. The colour scale gives an indication of the relative surface density. Right: nine sets of 300 points, randomly sampled from the corresponding fields on the left.}
\label{fig:FBM_cluster}
\end{figure*}

Here, we present a method for generating artificial star clusters from FBM density fields. \citet{PS88} provide multiple methods for generating the underlying field; we follow the spectral synthesis technique used by \citet{SBH98}. In addition, we define the BF and RDP cluster models used by CW04 to calibrate the $\mathcal{Q}$ estimator. These two models have a crossover point where they generate clusters with a uniform distribution. For a more in-depth discussion of the structural properties of the BF and RDP models, we refer the reader to CW04 and JWL17.

The generation of all three models relies heavily on pseudo-random number generation. Throughout this section, we define $\mathcal{U}$ as a random variate drawn from the uniform distribution in the interval $[0,1]$, and $\mathcal{G}$ as a variate from the Gaussian distribution with zero mean and unit variance. These models can be extended to any $E$-dimensional space. We use the shorthand $E_2$ and $E_3$ to indicate 2 and 3 dimensional space respectively.

\subsection{FBM clusters}
\label{sec:fbm_clusters}

We generate FBM clusters by generating an FBM probability density distribution. From this, we randomly sample $E$-dimensional variates, i.e. stellar positions. FBM is an $E$-dimensional generalisation of classical Brownian motion, parameterised by a drift exponent $H$ (sometimes referred to as the Hurst index), which may take a value between 0 and 1. The field's power spectrum is related to $H$ via the spectral index $\beta=E+2H$. For a 1-dimensional FBM curve $f(x)$, the value at $x+\Delta x$ is given by $f(x+\Delta x)=f(x)+\Delta f$, where $\Delta f$ is a random Gaussian increment. When $H=1/2$, i.e. classical Brownian motion, $\Delta f$ is uncorrelated with $f(x)$. When $H>1/2$, the curve is smoother, i.e. $\Delta f$ is correlated with $f(x)$. When $H<1/2$, the curve is rougher, i.e. $\Delta f$ is anticorrelated with $f(x)$. In $E$ dimensions, FBM structures have fractal dimension $D=E-H$. When $D$ is close to $E-1$, the structure is smooth and coherent (e.g. a single sheet, filament or core). When $D$ is close to $E$, the structure consists of multiple sub-clumps which are evenly distributed in space.

We generate the \changes{periodic} field $f(\boldsymbol{r},H)$ numerically on an $E$-dimensional Cartesian grid. Along each axis, $r$ has integer values in the range $1\leq r\leq N_\textsc{pix}$ (for $E_2$, we set $N_\textsc{pix}=1000$; for $E_3$ we set $N_\textsc{pix}=100$). First, we generate the spectrum,
\begin{equation}
	\hat{f}(\boldsymbol{k},H)=A(\boldsymbol{k},H)\left[\cos\varphi(\boldsymbol{k})+\mathrm{i}\sin\varphi(\boldsymbol{k})\right]\,,
\end{equation}
where $\boldsymbol{k}$ is a grid of wavevectors with integer $k$ values $\lfloor -N_\textsc{pix}/2 \rfloor\leq k\leq\lceil N_\textsc{pix}/2 \rceil$ along each axis. The amplitudes $A(\boldsymbol{k},H)$ and phases $\varphi(\boldsymbol{k})$ of each component of the spectrum are given by
\begin{equation}
	\begin{split}
		A(\boldsymbol{k},H)&=
			\begin{cases}
				\mathcal{P}^{-1/2}\,\lVert\boldsymbol{k}\rVert^{-\beta/2}&\text{if }\boldsymbol{k}\neq\boldsymbol{0}\,;\\
				0&\text{if }\boldsymbol{k}=\boldsymbol{0}\,,
			\end{cases}\\
		\mathcal{P}&=\sum\limits_{\boldsymbol{k}}\,\lVert\boldsymbol{k}\rVert^{-\beta}\,,\\
		\beta&=E+2H\,,
	\end{split}
	\label{eqn:amplitudes}
\end{equation}
and
\begin{equation}
	\begin{split}
		\varphi(\boldsymbol{k})&=\chi(\boldsymbol{k})-\chi(-\boldsymbol{k})\,,\\
		\chi(\boldsymbol{k})&=2\uppi\mathcal{U}\,.
	\end{split}
	\label{eqn:phases}
\end{equation}
The field $f(\boldsymbol{r},H)$ can be obtained by performing an inverse discrete Fourier transform on $\hat{f}(\boldsymbol{k},H)$.\footnotemark Note that the first line of Eqn. \ref{eqn:phases} ensures that $\hat{f}(-\boldsymbol{k},H)$ is the complex conjugate of $\hat{f}(\boldsymbol{k},H)$ and therefore $f(\boldsymbol{r},H)$ is strictly real.

\footnotetext{If $N_\textsc{pix}$ is even, the range of $k$-values along a given axis is reduced to $-N_\textsc{pix}/2\leq k\leq N_\textsc{pix}/2-1$. Here, the $N_\textsc{pix}/2$ wavenumber is equivalent to $-N_\textsc{pix}/2$. Values of  $\hat{f}(\boldsymbol{k},H)$ with one or more coordinates $k=N_\textsc{pix}/2$ are superposed onto the corresponding $-N_\textsc{pix}/2$ values.}

As noted by \citet{PS88} and JWL17, fractal structures in nature are self-similar over a limited range of length scales. It is therefore appropriate to introduce a length-scale $h$ at which the self-similarity of the structure ceases. This can be easily implemented by convolving $f(\boldsymbol{r},H)$ with a Gaussian kernel,
\begin{equation}
	\begin{split}
		f'(\boldsymbol{r},H,h)&= f(\boldsymbol{r},H)\ast w(\boldsymbol{r},h)\,,\\
		w(\boldsymbol{r},h)&=\frac{1}{h^E\,(2\uppi)^{E/2}}\exp\left(-\frac{\lVert\boldsymbol{r}\rVert^2}{2h^2}\right)\,.
	\end{split}
	\label{eqn:convolve}
\end{equation}
Here, $h$ is the smoothing length given in pixels widths. This is equivalent to applying a Gaussian filter to $\hat{f}(\boldsymbol{k},H)$ with standard deviation $k_\textsc{max}=N_\textsc{pix}/4h$.

The FBM field cannot directly be used as a PDF because, by construction, the distribution of $f'(\boldsymbol{r},H,h)$ is roughly Gaussian with $\langle f'(\boldsymbol{r},H,h)\rangle\approx0$ and $\langle f'(\boldsymbol{r},H,h)^2\rangle\approx1$. However the fractal properties of a structure remain unchanged when its density is transformed via a one-to-one function. Here, we exponentiate the field:
\begin{equation}
	g(\boldsymbol{r},H,h,\sigma)=\exp\left[\frac{\sigma f'(\boldsymbol{r},H,h)}{\sqrt{\left\langle f'(\boldsymbol{r},H,h)^2\right\rangle}}\right]\,,
\end{equation}
where $\sigma$ is a free parameter. This changes the Gaussian distribution of densities into a lognormal distribution. Note that $\sigma$ is the standard deviation of the natural log of $g(\boldsymbol{r},H,h,\sigma)$.
 
Finally, We circularly shift the FBM field so that its periodic centre of mass lies at the centre of the grid. This tends to place coherent structures within high $H$ fields at the centre and lower density regions around the edges. 

In summary, we generate a modified FBM field, defined using three parameters: $H$, $h$ and $\sigma$.\footnotemark This is then used as the PDF from which we sample $N_\star$ random positions (see Appendix \ref{apn:random} for a description of the random sampling technique). $E_3$ clusters are projected onto $E_2$ space by marginalising the distribution along one of its axes. We note that in most practical cases (i.e. $N_\star\leq10^4 $), $h$ is unlikely to have a strong impact on the distribution of points. Essentially, $h$ is a \emph{nuisance parameter} which we include to randomise the field resolution without introducing coarse grid artefacts. For $E_2$ fields, we randomly pick a value of $h$ from the log-uniform distribution in the interval \mbox{$[10^{-3}\,N_\textsc{pix},10^{-2}\,N_\textsc{pix}]$}. For $E_3$ fields, computational limitations require we use a coarser grid (both grids have the same number of elements). Here, we skip Eqn. \ref{eqn:convolve} and set $f'(\boldsymbol{r},H,h)=f(\boldsymbol{r},H)$.

Fig. \ref{fig:FBM_cluster} shows how the structure of an $E_2$ FBM cluster varies with $H$ and $\sigma$. Here, we have used the same random seed for each realisation and set $h=10^{-3} N_\textsc{pix}$. We see that fixing $\sigma$ and varying $H$ alters the amount of substructure in a cluster. The outline of the cluster remains roughly the same shape, but the number of internal clumps increases with $H$. Fixing $H$ and changing $\sigma$ alters the dynamic range of the cluster surface-density. When $\sigma$ is high, the clumps are sharply defined. As $\sigma$ tends towards zero, the cluster structure tends towards uniform density distribution.

\footnotetext{Strictly speaking, the field is defined by five parameters, if we include $N_\textsc{pix}$ and the random seed.}

\subsection{Box-fractal cluster}

We generate a BF cluster with approximately $N_\star$ stars by taking an $E_3$ cube with unit edge-length, and bisecting it along each axis to make $2^E$ sub-cubes. A random set of $2^D$ sub-cubes are labeled as \emph{active}, where $D$ has a value in the interval $(0,E]$. In cases where $2^D$ is non integer, the number of active sub-cubes if given by,
\begin{equation}
{\setstretch{1.4}
	N_\textsc{act}=
		\left\{\begin{array}{cc}
			\floor*{2^D}& \text{if } \left(2^D-\floor*{2^D}\right) < \mathcal{U}\,;\\
			\ceil*{2^D}& \text{if } \left(2^D-\floor*{2^D}\right)\geq\mathcal{U}\,.\\
		\end{array}\right.
}
\end{equation}
The method is recursively repeated on each active sub-cube a further $\ceil*{\log_2(N_\star)/D}-1$ times. Finally, a star is placed at a random position within the volume of each final generation cube. In order to perform the analysis in $E_2$ space, the BF structure is projected through a random line of sight.

\subsection{Radial profile cluster}
We construct an $E_3$ RDP cluster by generating $N_\star$ random coordinates,
\begin{equation}
	\begin{split}
		\boldsymbol{r}&=r\,\boldsymbol{\hat{u}}\,,\\
		r&=\mathcal{U}^\frac{1}{E-\alpha}\,,\\
		\boldsymbol{u}&=(\mathcal{G}_1,\mathcal{G}_2,\ldots,\mathcal{G}_E)\,,
	\end{split}
	\label{eqn:radial_cluster}
\end{equation}
where $\alpha$ is the radial density exponent. The cluster has a density profile $\rho(r)\propto r^{-\alpha}$, where $\alpha$ may have any value in the interval $[0,E)$. Again, the $E_3$ cluster is projected onto $E_2$ space through a random line of sight.

\section{Parameter estimators}
\label{sec:parameter_estimators}

\begin{table}
	\centering
	\begin{tabularx}{\columnwidth}{ccccX}
		\hline
		\# & Name & $N_\star$ & $D\,\mathrm{[pc]}$ & Reference \\ 
		\hline
		1 & Lupus 3 & 67 & 170 & \citet{C08} \\
		2 & IC 348 & 350 & 315 & \citet{LML06,MLL07} \\
		3 & $\rho$ Oph & 198 & 130 & \citet{BAK01} \\
		4 & IC 2391 & 200 & 150 & \citet{BSB01} \\
		5 & Cha I & 234 & 160 & \citet{L07} \\
		6 & Taurus & 335 & 140 & \citet{LAE10} \\
		\hline
	\end{tabularx}
	\caption{Cluster properties and sources. The first column gives the numeric identifier used throughout this paper; the second column gives the name of the cluster; the third column gives the number of stellar objects (after multiple systems are fused into single objects); the fourth column gives the assumed distance to the cluster; the fifth column cites the source of the data.}
	\label{tab:clusters}
\end{table}

\begin{figure*}
\centering
\includegraphics[width=0.85\columnwidth]{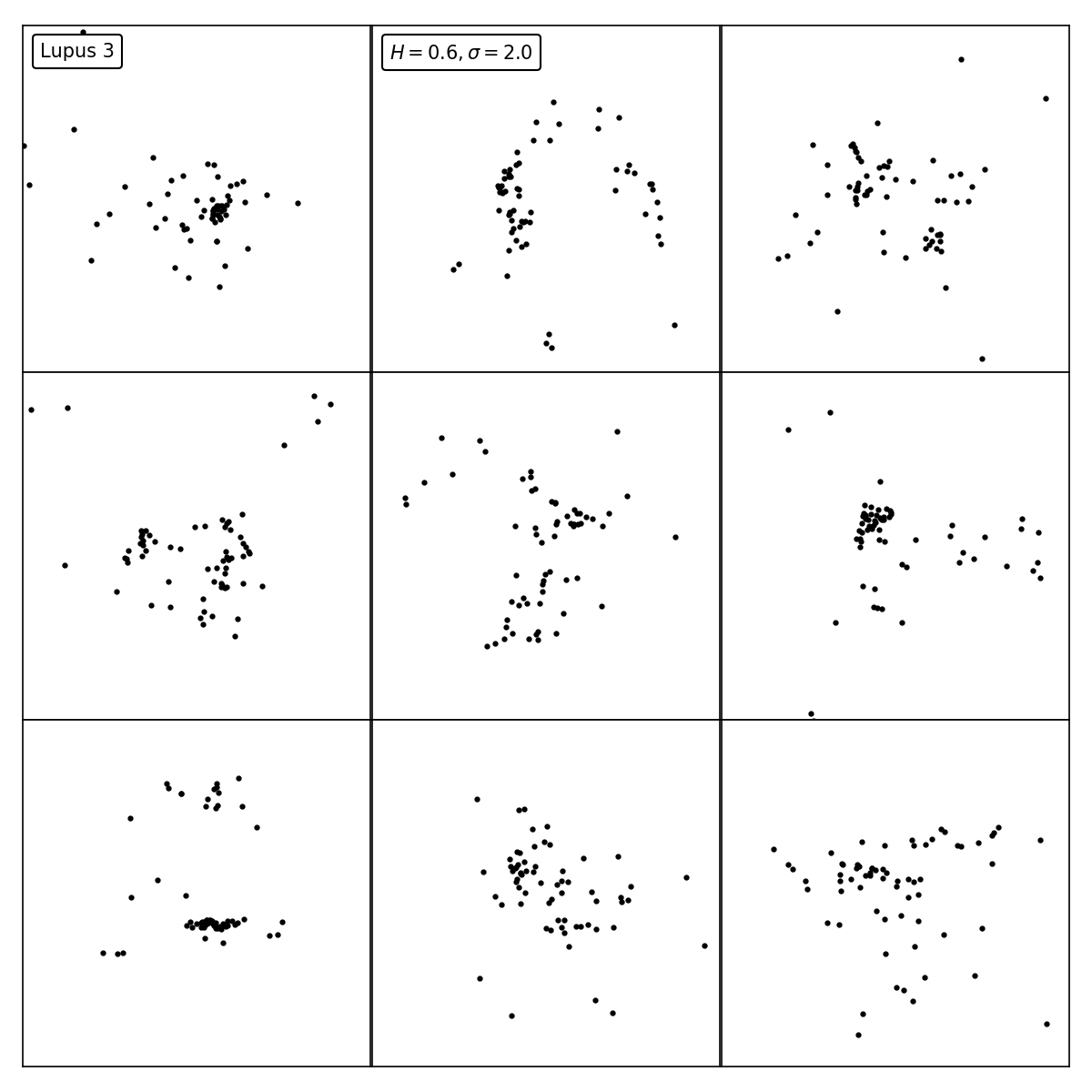}
\includegraphics[width=0.85\columnwidth]{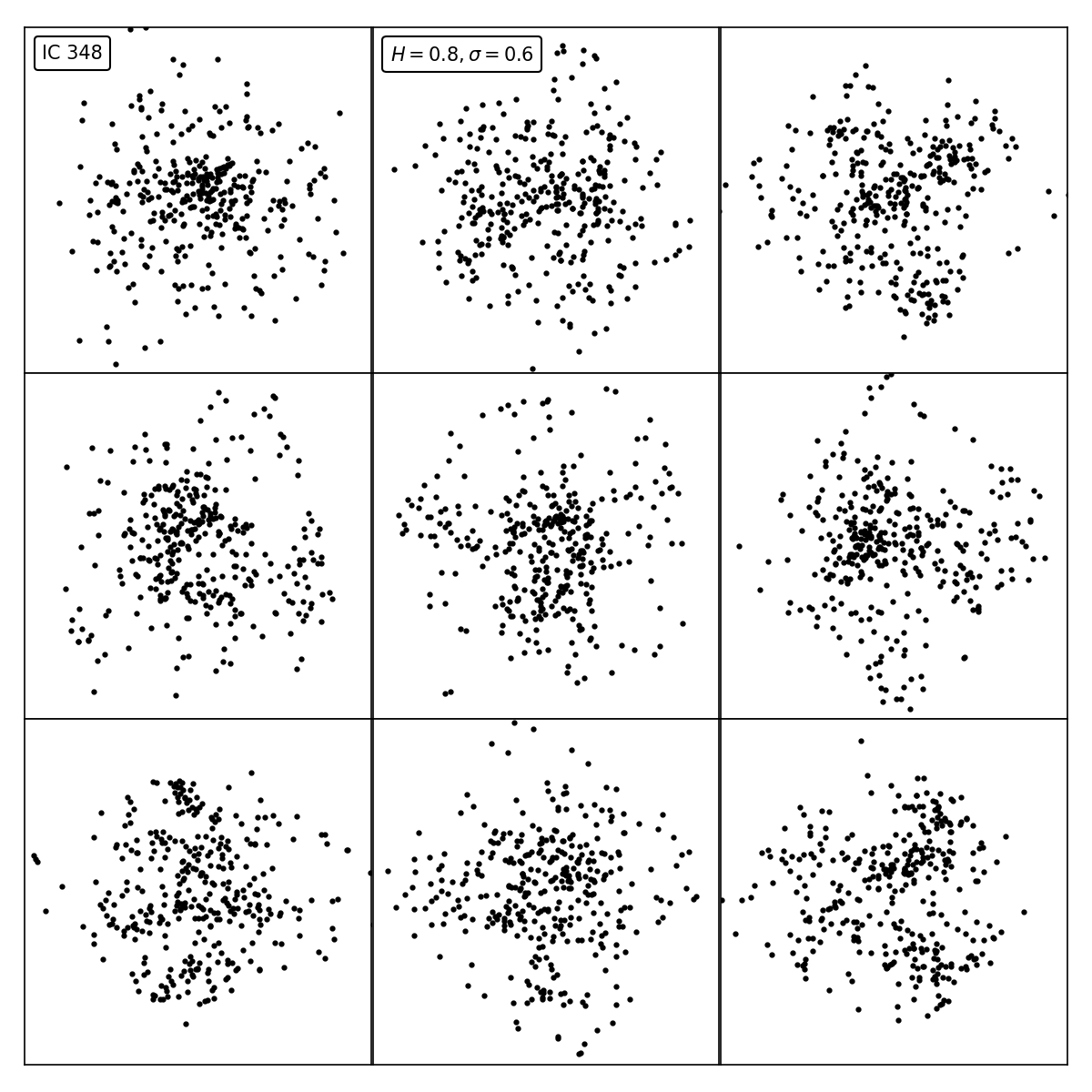}
\includegraphics[width=0.85\columnwidth]{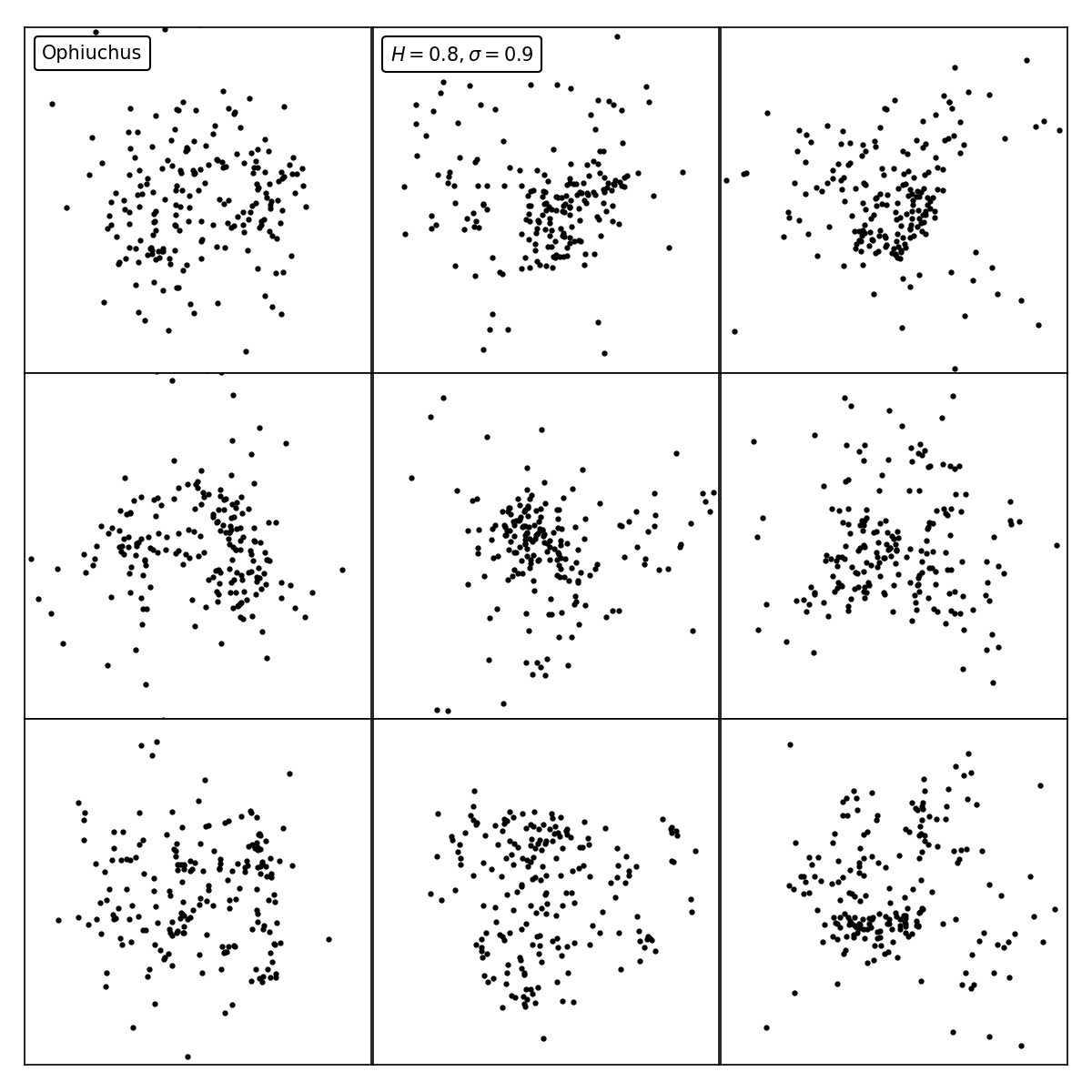}
\includegraphics[width=0.85\columnwidth]{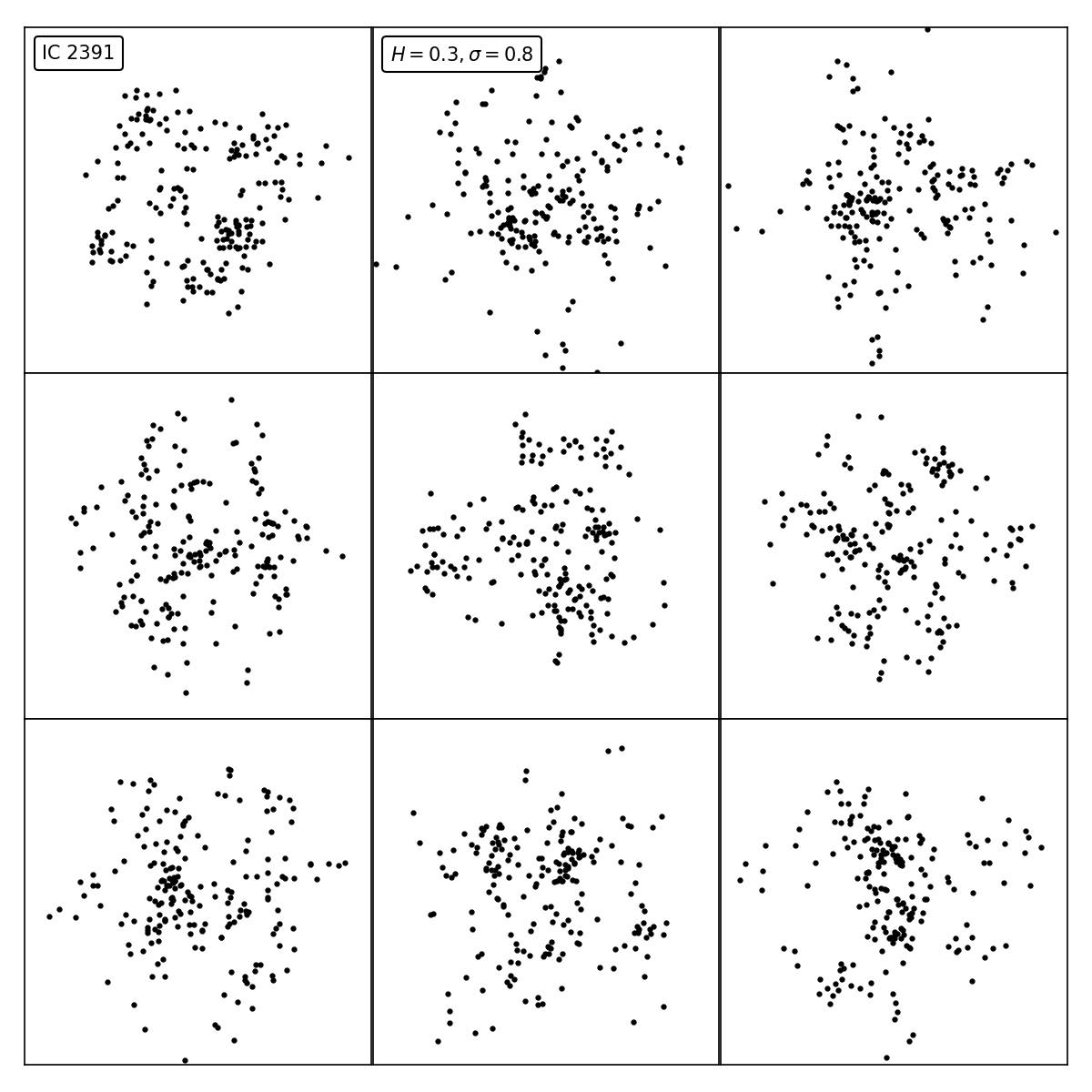}
\includegraphics[width=0.85\columnwidth]{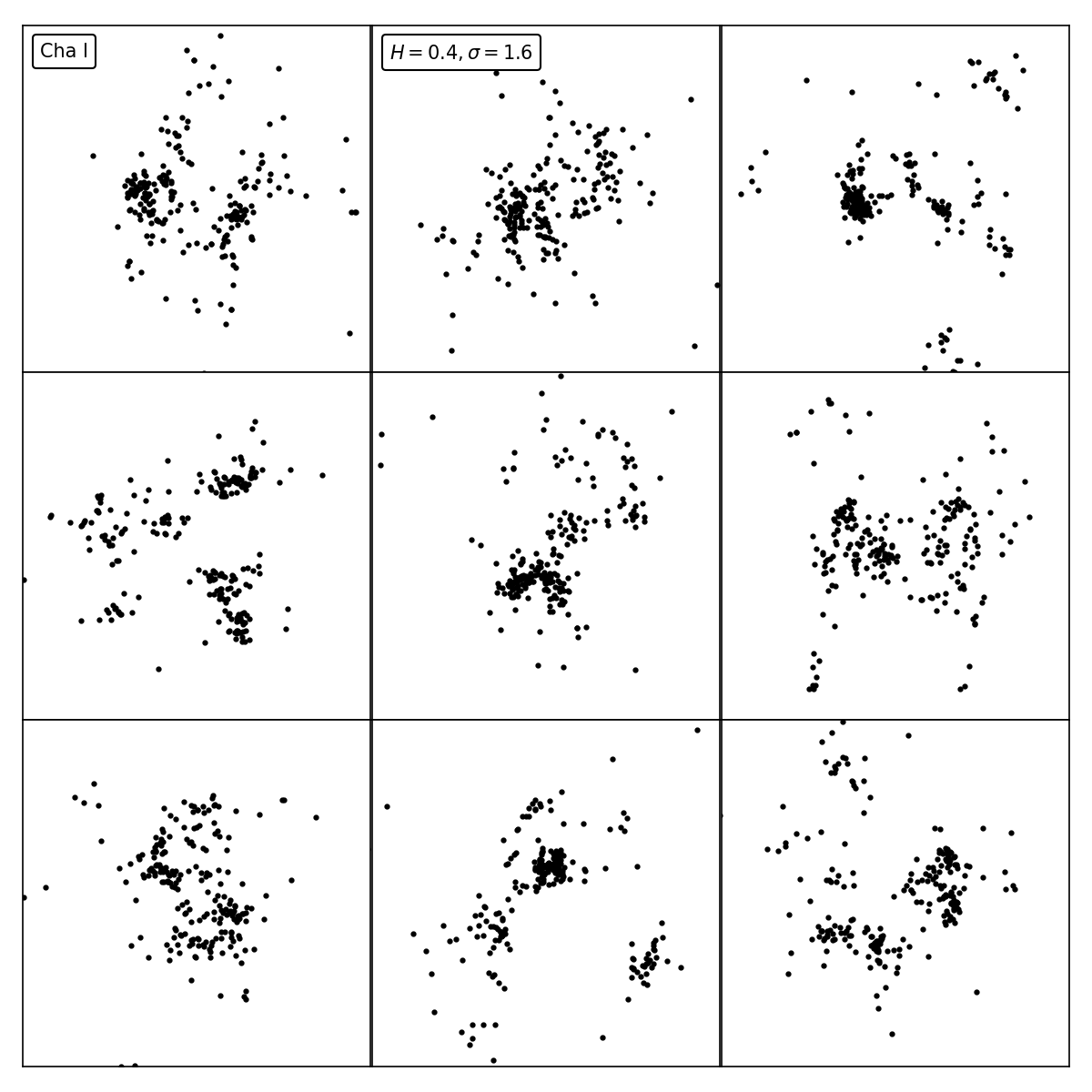}
\includegraphics[width=0.85\columnwidth]{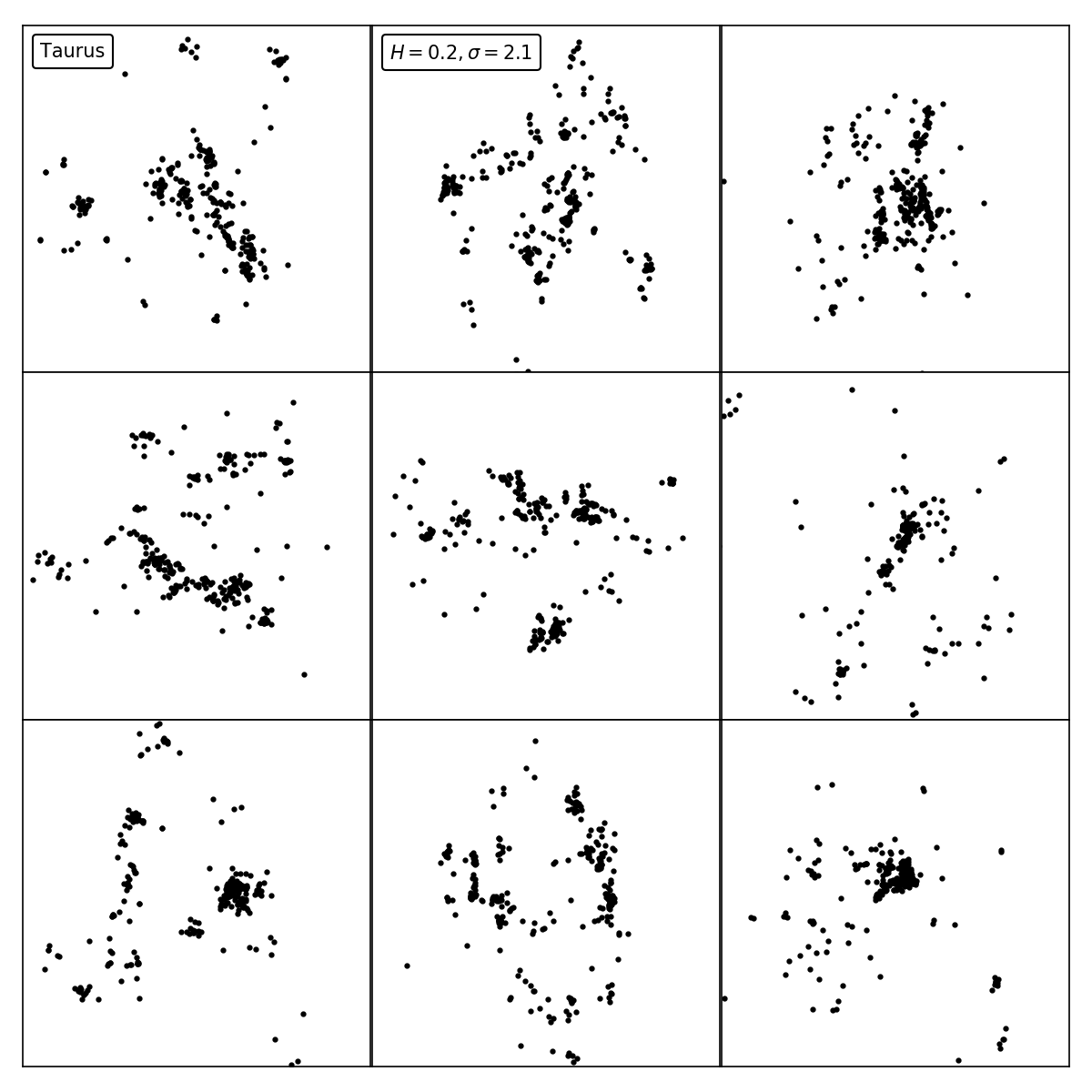}
\caption{The positions of stars in real clusters, presented alongside those of artificial FBM clusters. In each of the six frames, the real cluster is plotted in the top-left corder. The remaining eight frames show different realisations of FBM clusters with the most likely estimated parameters (see Section \ref{sec:machine_learning}). In all cases, the aspect ratios of the clusters have been removed (see Eqn. \ref{eqn:whitening}).}
\label{fig:clusters}
\end{figure*}

We examine a family of estimators which use the minimum spanning tree (MST) and complete graph (CG) to infer the structural parameters of clusters. The MST is the shortest possible network which connects $N_\star$ vertices with \mbox{$N_m=N_\star-1$} edges. The CG is the graph which connects each vertex directly to all the other vertices. The CG has \mbox{$N_s=N_\star(N_\star-1)/2$} edges in total. \changes{Here, we review the $\mathcal{Q}$ estimator (CW04) and $\bar{m}$-$\bar{s}$ plots \citep[][hereafter C09]{C09a}. Next, we present a machine learning algorithm which builds and improves upon these two methods. In all three cases, we (i) test each estimator's ability to recover the parameters of artificial clusters, and (ii) apply them to a selection of observed clusters.}

\subsection{Observations}

\changes{\noindent{}We apply the estimators to the clusters examined by CW04 and JWL17. Table \ref{tab:clusters} lists their properties and original references. The stellar positions of each cluster are plotted in Fig. \ref{fig:clusters}. Each of cluster has been preprocessed to remove probable multiple systems. Here, any star with neighbours closer than $5\times10^{-3}\,\mathrm{pc}$ is removed, along with its neighbours, and replaced by a single star at the original stars' centre of mass. This minimum length scale reflects the widest separations typically observed amongst multiple systems in young clusters \citep{KPPG12}.}

\subsection{$\mathcal{Q}$ parameter}

\begin{figure*}
\centering
\includegraphics[width=\textwidth]{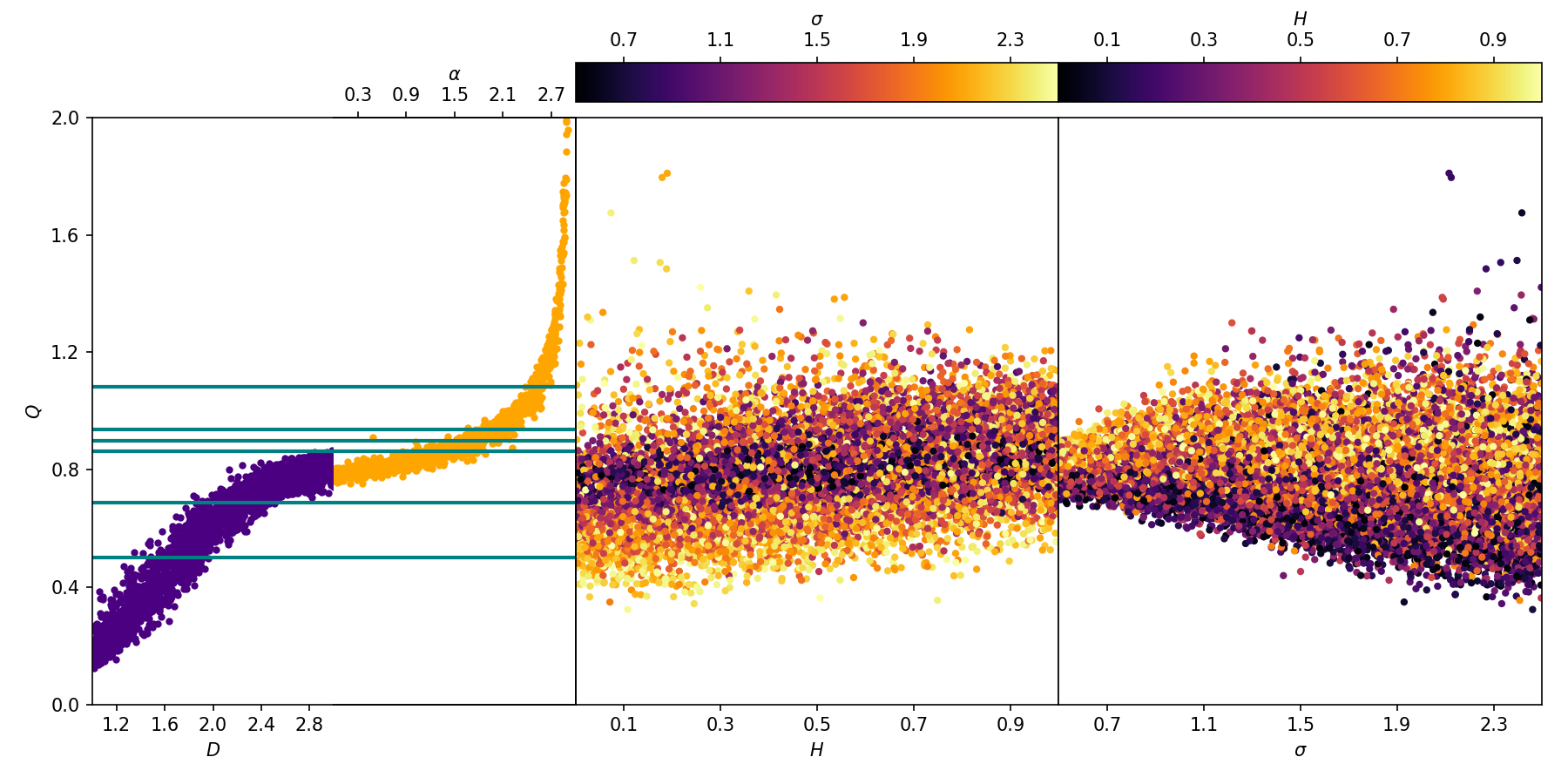}
\caption{$\mathcal{Q}$ values plotted against the underlying parameters of artificial star clusters. Each artificial cluster contains between 300 and 1000 stars. The left panel shows $\mathcal{Q}$ values for BF clusters (purple) and RDP clusters (gold). Each set of points represents 2000 clusters with random $D$ in the interval $[1,3]$ and $\alpha$ in the interval $[0,3)$. The horizontal teal lines show the values of $\mathcal{Q}$ calculated for the clusters listed in Table \ref{tab:clusters} (the vertical order of the lines is the same as the table). The centre and right panels show the $\mathcal{Q}$ values for FBM clusters as a function of $H$ and $\sigma$. Each panel contains 5000 clusters with $H$ in the interval $[0,1]$ and $\sigma$ in the interval $[0.5,3.5]$. In each case, the colour scale gives the value of the other parameter.}
\label{fig:Q_plot}
\end{figure*}

CW04 define the statistic $\mathcal{Q}=\bar{m}/\bar{s}$, where $\bar{m}$ and $\bar{s}$ are respectively the normalised mean edge lengths of the MST and the CG:
\begin{equation}
	\begin{split}
		\bar{m}&=\left(\frac{1}{\left(\uppi R^E[N_m+1]^{E-1}\right)^{1/E}}\right)\,\sum\limits_{i=1}^{N_m}m_i\,,\\
		\bar{s}&=\left(\frac{1}{N_sR}\right)\,\sum\limits_{i=1}^{N_s}s_i\,.
	\end{split}
	\label{eqn:mbar_sbar}
\end{equation}
Here, $m_i$ and $s_i$ are graph edge-lengths and $R$ is a characteristic length-scale of the system. Note that the $R$-terms cancel when calculating $\mathcal{Q}$.

The \changes{CW04} calibration of $\mathcal{Q}$ involves calculating the statistic for BF and RDP clusters. A uniform density cluster (i.e. $D=3$ or $\alpha=0$) returns $\mathcal{Q}\sim0.8$. BF clusters have $\mathcal{Q}\lesssim0.8$ and RDP clusters have $\mathcal{Q}\gtrsim0.8$. $\mathcal{Q}$ increases monotonically with both $D$ and $\alpha$.  Fig. \ref{fig:Q_plot} shows the relationship between $\mathcal{Q}$ and $D$, and between $\mathcal{Q}$ and $\alpha$.

The $\mathcal{Q}$ values of Lupus 3, IC 348 and $\rho$ Oph suggest that they have radial density profiles with $\alpha\gtrsim1.5$. The $\mathcal{Q}$ values of Cha I and Taurus suggest they are similar to BF structures with $D\lesssim2.5$. The $\mathcal{Q}$ value of IC 2391 lies near a plateau on the plot, making its structural type difficult to determine.

Fig. \ref{fig:Q_plot} also shows how $\mathcal{Q}$ relates to the parameters of FBM clusters. Here, we see that there is a slight positive correlation between between $H$ and $\mathcal{Q}$. However, the scatter introduced by $\sigma$ exceeds the dynamic range of the correlation. There is no noticeable correlation between $\mathcal{Q}$ and $\sigma$.  Therefore $\mathcal{Q}$ is a poor predictor of $H$ and/or $\sigma$.

\subsection{$\bar{m}$-$\bar{s}$ plots}

C09 suggest that plots of $\bar{m}$ versus $\bar{s}$ provide a more robust diagnostic tool than $\mathcal{Q}$ alone. They show that BF and RDP clusters with fixed parameters fill distinct regions of the $\bar{m}$-$\bar{s}$ plot. However, there is a lack of agreement on which length scale $R$ should be used to normalise $\bar{m}$ and $\bar{s}$. In the original \changes{C09} publication, $R$ is set to the distance between the cluster's centre of mass and its outer most point. This measure is problematic as (i) a single outlying star can dominate the length-scale and (ii) this value is not representative of the area of a cluster with a high aspect ratio. Both of these issues can add significant noise to the normalisation of $\bar{m}$ and $\bar{s}$ \citep[see][for a review of different $R$ normalisation methods]{P18}. Instead, we use the \citet{SK06} scheme which sets $R$ to the square root of the area of the convex hull of the set of stars. This lessens (although does not necessarily eliminate) the issues with outliers and the aspect ratio.

\begin{figure}
\centering
\includegraphics[width=\columnwidth]{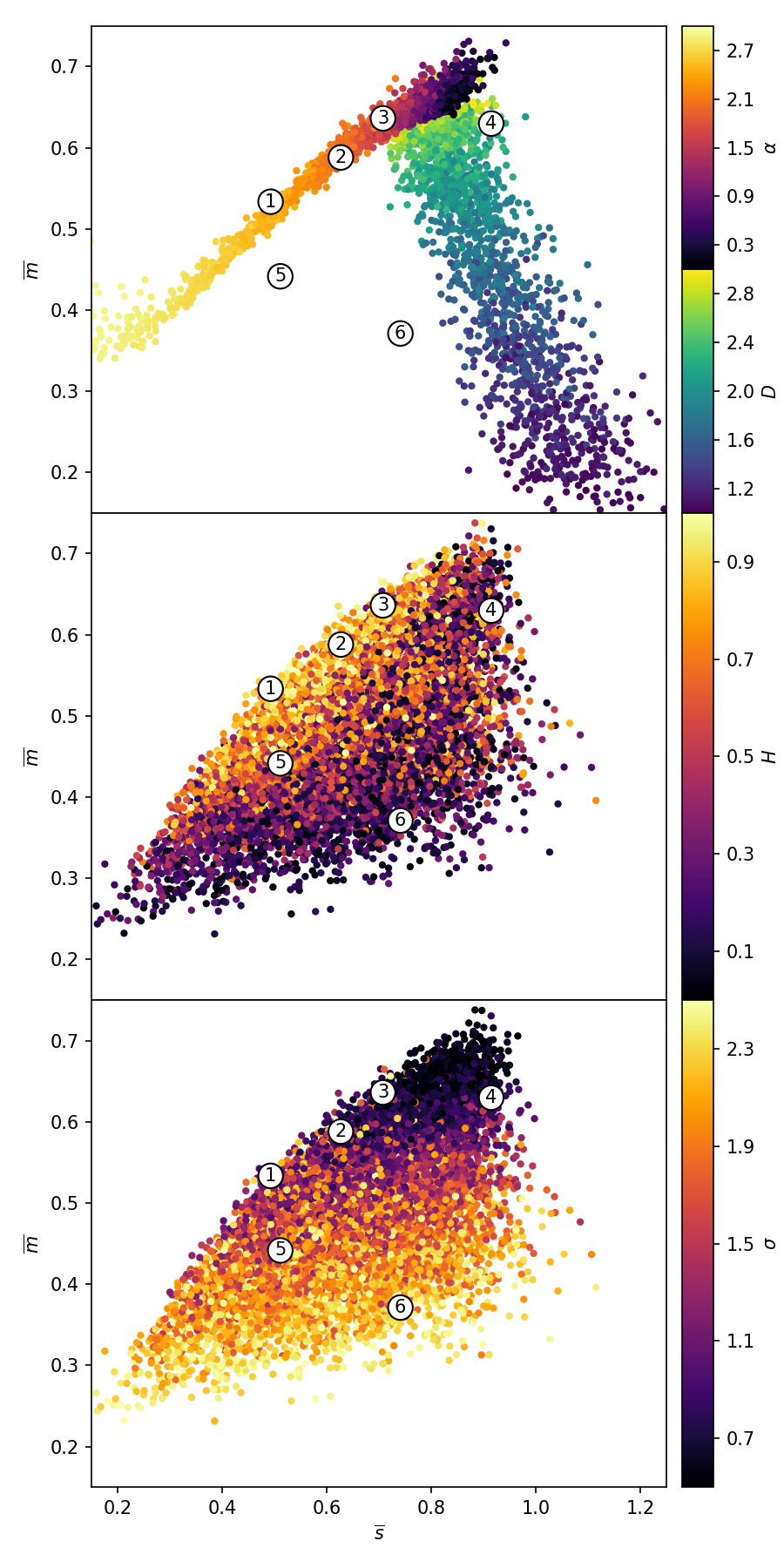}
\caption{Plots of $\bar{m}$ versus $\bar{s}$ for the BF and RDP clusters (top panel) and the FBM clusters (middle and lower panel). The colour scale gives the values of the underlying cluster parameters. The points are generated from the same clusters as Fig. \ref{fig:Q_plot}. The numbered points give $\bar{m}$-$\bar{s}$ values for the clusters given in Table \ref{tab:clusters}. In all cases, the value of $R$ (see Eqn. \ref{eqn:mbar_sbar}) is set to the square root of the area of the cluster convex hull.}
\label{fig:mbar_sbar_plot}
\end{figure}

The top frame of Fig. \ref{fig:mbar_sbar_plot} shows how model clusters with different values of $D$ or $\alpha$ occupy different regions on the $\bar{m}$-$\bar{s}$ plot. Here, the parameter estimates for Lupus 3, IC 348 and $\rho$ Oph are unchanged from their respective $\mathcal{Q}$ estimates. In addition, the plot suggests that IC 2391 is similar to a BF cluster with $D\sim2.8$. However, we find that  Taurus and Cha I do not match up to \emph{any} of the BF or RDP clusters. On visual inspection (see Fig. \ref{fig:clusters}), they are clearly sub-clustered, but their $\bar{m}$-$\bar{s}$ values cannot be matched to any value of $D$.

The remaining two frames of Fig. \ref{fig:mbar_sbar_plot} shows the $\bar{m}$-$\bar{s}$ values for FBM clusters. Here, unlike the BF and RDP models, the FBM clusters fill an area of the plot which overlaps all of the observed clusters. We see that FBM clusters with $H\sim1$ fill the same region of the plot as RDP clusters. This is unsurprising, as they both represent smoothly distributed, centrally concentrated clusters. However, clusters with $H\lesssim1$ do not appear to occupy distinct regions of the plot. Finally, We see a very strong negative correlation between $\bar{m}$ and $\sigma$. This shows that the mean edge-length of the minimum spanning tree is much more sensitive to the surface-density dynamic range of a cluster than its fractal properties.

\subsection{Machine learning regression}
\label{sec:machine_learning}

\begin{figure*}
\centering
\includegraphics[width=\textwidth]{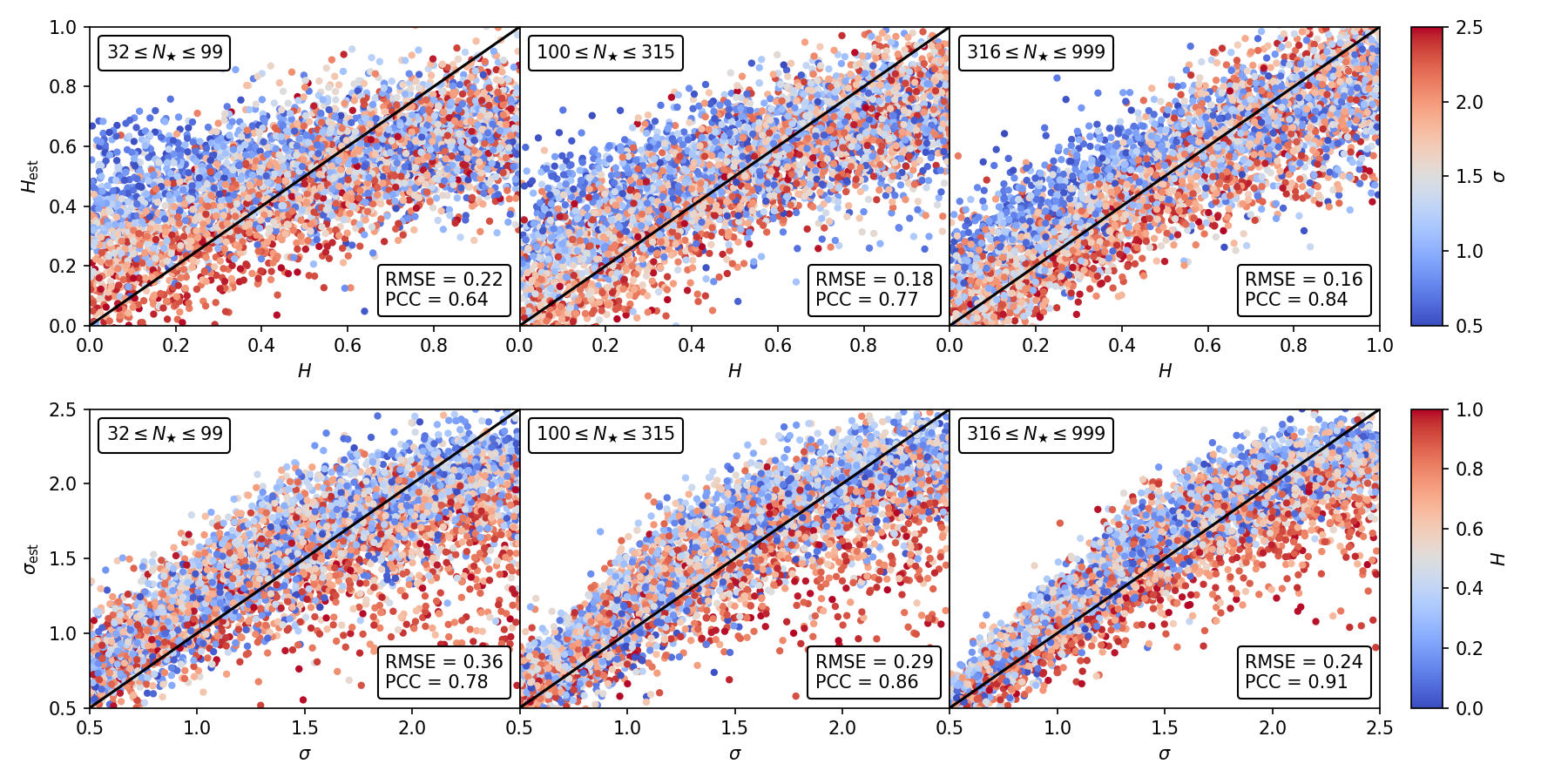}
\caption{Test data parameter estimates as a function of the underlying parameters. The top row shows the ANN's ability to predict $H$. The bottom row shows ANN's ability to predict $\sigma$. The colour scale gives the value of the other parameter. The range of $N_\star$ is indicated in the top-left corner of each plot. The solid black line shows the hypothetical performance of a perfect estimator. We also give the root-mean-squared error and Pearson's correlation coefficient for each plot in the bottom-right corner.}
\label{fig:uncertainties}
\end{figure*}

\begin{figure}
\centering
\includegraphics[width=\columnwidth]{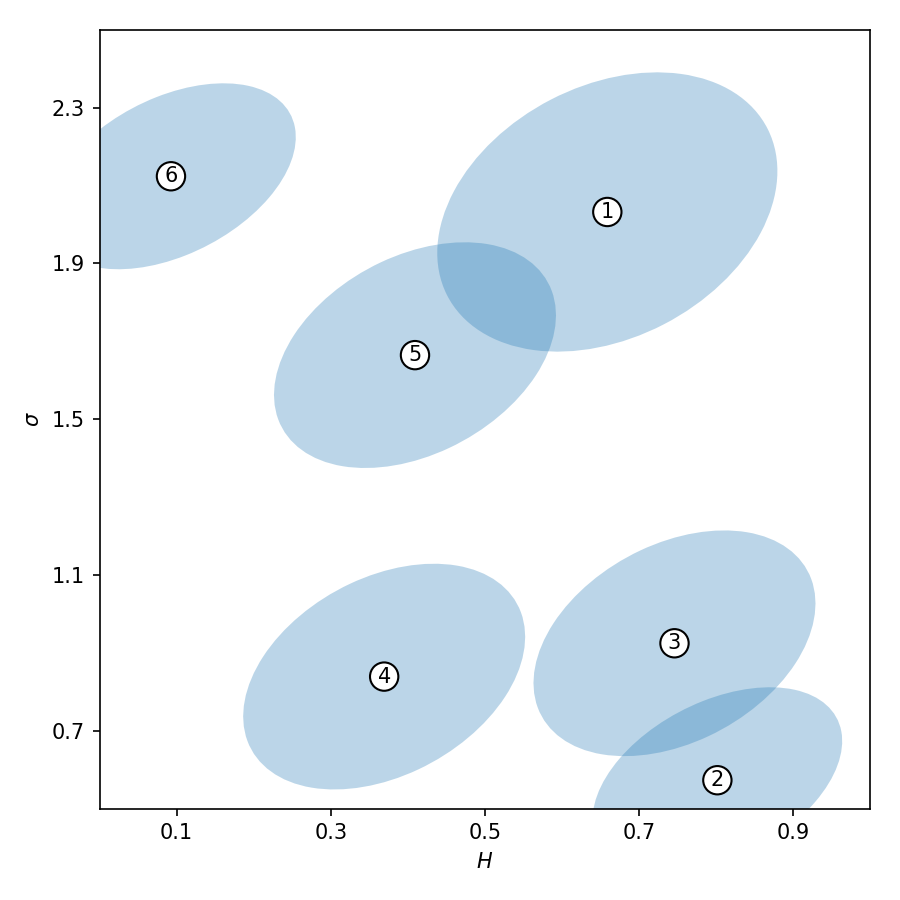}
\caption{A scatter plot of the $E_2$ estimates of $H$ and $\sigma$ for the clusters listed in Table \ref{tab:clusters}. The ellipses show the one-sigma uncertainties, calculated from the root-mean-squared errors and their covariance.}
\label{fig:H_sigma_plot}
\end{figure}

\begin{table}
	\centering
	\begin{tabular}{ccc}
		\hline
		Parameter & Distribution & Interval \\
		\hline
		$N_\star$ & log-uniform & $[32,99]$, $[100,315]$, $[316,999]$ \\
		$H$ & uniform & $[0,1]$ \\
		$\sigma$ & uniform & $[0.5,3.5]$ ($E_2$), $[0.5,4.5]$ ($E_3$) \\
		\hline
	\end{tabular}
	\caption{The range and distributions of parameters used to train the ANN regressor. The first column gives the parameter; the second column gives the type of distribution; the third column gives the interval. Note that $\sigma$ has a different ranges for $E_2$ and $E_3$.}
	\label{tab:param_range}
\end{table}

$\mathcal{Q}$ and $\bar{m}$-$\bar{s}$ plots are often used to estimate underlying parameters by visual inspection. \changes{By this, we mean that a large ensemble of $\mathcal{Q}$ or $\bar{m}$-$\bar{s}$ measurements for a known set of models are plotted; parameters are attributed to an observation based on the plot-distance from the observation's measurements to the equivalent model values.} This methodology makes it difficult to quantify parameter uncertainties. Furthermore, we have shown that the BF model, which typically is used to calibrate the two methods, is unable to produce clusters with similar properties to Taurus or Cha I. The latter of these two problems may be addressed by implementing the FBM cluster model. However, the $\mathcal{Q}$ and $\bar{m}$-$\bar{s}$ methods are poor at distinguishing the underlying parameters. We address these shortcomings with a machine learning regressor which uses FBM clusters as training data.

\changes{
\noindent A \emph{regressor} is an analytical function or numerical procedure $F(x)$ which gives an estimate of $y$ for a given input (or feature) $x$. In order to make these estimates, the regressor must first be \emph{trained}. A simple example of a regressor is linear regression, i.e. $F(x)=m\,x+c$. Training the regressor involves taking $N$ training data, $x_i$ and $y_i$, and finding values $m$ and $c$ (hyperparameters\footnotemark) which minimise a loss statistic, e.g. $L=\sum_{i=1}^N (F(x_i)-y_i)^2$.
}

\footnotetext{In most contexts, these are referred to as \emph{parameters}. We refer to them as \emph{hyperparameters} so that they are not confused with cluster model parameters, e.g. $H$ and $\sigma$.}

\changes{
A similar approach can be used to estimate the parameters of a star cluster. Here, $\boldsymbol{x}$ is a vector of statistics that are directly taken from the cluster (we define these in Section \ref{sec:training}), and $\boldsymbol{y}=(H,\sigma)$ is the vector of underlying parameters. Here, the regressor $F(\boldsymbol{x})$ is an Artificial Neural Network (ANN). Complex ANNs are routinely used in fields such as image analysis \citep[e.g.][]{LBBH98}. However, comparatively simple ANNs can be used for numerical regression problems with multiple inputs and outputs \citep[e.g.][]{RAD18}.}

\changes{
Details of ANN used here, along with links to the full implementation in $\textsc{Python}$, are given in Appendix \ref{sec:ANN}. If the reader is not concerned about these technical details, they should simply note that the ANN hyperparameters are estimated from \emph{training data}. Once trained, the ANN is applied to \emph{test data}. This enables us to (i) ensure we are not overfitting the training data and (ii) quantify the uncertainties of the regressor. In the following sections, we discuss the training, testing and the results of the ANN.}

\subsubsection{Training}
\label{sec:training}

For each star cluster, we generate a set features $\boldsymbol{x}$ using its CG and MST. However, as noted by \citet{CW09}, the elongation of a star cluster may affect these graphs. Before we build the graphs, we \emph{whiten} the distribution of points. This completely removes the size scale and aspect ratio from the distribution. We calculate the covariance matrix $\varSigma(\boldsymbol{r})$ for the set of stellar positions $\boldsymbol{r}$. From this, we calculate a new set of positions $\boldsymbol{r}'$, where each value $\boldsymbol{r}'_i$ has elements,
\begin{equation}
	\begin{split}
		r'_{ij}&=\frac{1}{\sqrt{\lambda_j}}\,\boldsymbol{r}_i\cdot\hat{\boldsymbol{v}}_j\,,\\
		j&=1,2,\ldots,E\,.
	\end{split}
	\label{eqn:whitening}
\end{equation}
Here, $\lambda_j$ and $\boldsymbol{v}_j$ are respectively the $j$th eigenvalues and eigenvectors of $\varSigma(\boldsymbol{r})$. Note that $\varSigma(\boldsymbol{r}')$ is equal to the identity matrix $\boldsymbol{I}$.

For a set of $N$ graph edges $l$, we define the mean edge-length $\mu(l)$, and the $n$th central moment $M_n(l)$, as
\begin{equation}
	\begin{split}
		\mu(l)&=\frac{A}{N}\sum\limits_{i=1}^N l\,,\\
		M_n(l)&=\left( \frac{A}{N} \sum\limits_{i=1}^N \left[l_i-\frac{\mu(l)}{A}\right]^N \right)^\frac{1}{N}\,,\\
		A&=\left\{
		\begin{array}{ll}
			\frac{N}{(N+1)^{(E-1)/E}} & \text{for MST;} \\
			1 & \text{for CG.}
		\end{array}\right.
	\end{split}
	\label{eqn:graph_moments}
\end{equation}
We construct $\boldsymbol{x}$ using the mean and the second, third and fourth central moments of the MST and CG edge-lengths. Note that the second, third and fourth central moments are related to the variance, skewness and kurtosis. We do not need to normalise these features to a length scale as we have already whitened the distribution of points.

We perform two analyses; one with $E_2$ FBM clusters and another with $E_3$. For each analysis, we train three regressors with different ranges of $N_\star$. \changes{There are two reasons for this. First, it is useful to quantify parameters uncertainties as a coarse function of $N_\star$. Second, the MST normalisation (see Eqn. \ref{eqn:graph_moments}) is technically only valid for the limit $N_\star\to\infty$ \citep{SM88}. Splitting the analysis into different $N_\star$ bins helps to isolate any biases which may occur as a function of $N_\star$.} For each regressor, we generate $10^4$ training clusters with randomly sampled values of $N_\star$, $H$ and $\sigma$. The ranges and distributions of these parameters are given in Table \ref{tab:param_range}.

\subsubsection{Testing and results}

We test each trained ANN by generating an additional $5\times10^3$ artificial clusters. These are randomly generated the same way as the training clusters, but with different random seeds. Fig. \ref{fig:uncertainties} shows the estimated parameters of $E_2$ test clusters as a function of their underlying true parameters. From these plots, we see that the parameters can be estimated with a useful degree of accuracy for clusters with $N_\star>100$. We can approximate the uncertainties as the root-mean-squared errors, $\Delta H=\sqrt{\langle(H_\textsc{est}-H)^2\rangle}$ and $\Delta \sigma=\sqrt{\langle(\sigma_\textsc{est}-\sigma)^2\rangle}$. Here, the \textsc{est} subscripts denote estimated parameters; the terms with no subscripts denote underlying parameters. For both the $E_2$ and $E_3$ cases, $\Delta H\approx\pm0.2$. For the $E_2$ case, $\Delta \sigma$ varies from $\pm0.3$ to $\pm0.5$. For $E_3$, $\Delta \sigma$ varies between $\pm0.5$ and $\pm0.7$. The magnitudes of the uncertainties decrease as $N_\star$ increases (we give values for $E_2$ clusters to two significant figures in Fig \ref{fig:uncertainties}). We also find that the $H$ and $\sigma$ uncertainties are correlated, i.e. there is some degeneracy in the expression of the two parameters. Here, high $\sigma$ can make a smooth distribution (determined by $H$) appear rougher, and \emph{vice versa}. We note that this uncertainty approximation may underestimate the error on $H_\textsc{est}$ when $N_\star<100$. Here, the correlation between $H_\textsc{est}$ and $H$ is visibly less tight than the other cases. 

Table \ref{tab:cluster_results} shows the parameter estimates for the observed star clusters. We find that for these six cases, $H$ appears invariant with respect to $E$, whereas the $E_3$ values of $\sigma$ are approximately one and a half times greater than the $E_2$ values. We also include approximate $D$ and $\alpha$ values estimated using $\bar{m}$-$\bar{s}$ plots for comparison. Fig. \ref{fig:H_sigma_plot} shows a plot of $\sigma$ against $H$ for the $E_2$ analysis. Here, we see that Taurus, Cha I and IC 2391 have similar levels of fractal structure to each other (determined by $H$), but are distinguished by different surface-density dynamic ranges (determined by $\sigma$). IC 348 and $\rho$ Oph are indistinguishable from one another, each with a smooth structure and a low level of surface-density variation. Lupus 3 has a high amount of surface-density variation, but the relatively low number of stars makes it difficult to estimate the uncertainty on $H$.

\begin{table*}
	\begin{tabulary}{\textwidth}{CCCCCCCC}
		\hline
		\# & Cluster & $H\,(E_2)$ & $\sigma\,(E_2)$ & $H\,(E_3)$ & $\sigma\,(E_3)$ & $D$ (C09) & $\alpha$ (C09) \\ 
		\hline
		1 & Lupus 3 & $0.6\pm0.2$ & $3.0\pm0.5$ & $0.6\pm0.2$ & $3.8\pm0.7$ & -- & $\sim2.5$ \\
	        2 & IC 348 & $0.8\pm0.2$ & $1.2\pm0.3$ & $0.7\pm0.2$ & $1.6\pm0.5$ & -- & $\sim2.1$ \\
	        3 & $\rho$ Oph & $0.7\pm0.2$ & $1.4\pm0.4$ & $0.7\pm0.2$ & $2.1\pm0.6$ & -- & $\sim1.8$ \\
	        4 & IC 2391 & $0.3\pm0.2$ & $1.3\pm0.4$ & $0.3\pm0.2$ & $2.0\pm0.6$ & $\sim2.8$ & -- \\
	        5 & Cha I & $0.4\pm0.2$ & $2.7\pm0.4$ & $0.2\pm0.3$& $3.7\pm0.6$ & -- & -- \\
	        6 & Taurus & $0.0\pm0.2$ & $3.2\pm0.3$ & $0.0\pm0.2$ & $4.7\pm0.5$ & -- & -- \\
		\hline
	\end{tabulary}
	\caption{Parameter values estimated for observed star clusters. The first and second columns give the identifier and name of the cluster. The third and forth columns give the $E_2$ FBM parameters, inferred using the ANN. The fifth and sixth columns give the same values for $E_3$. The seventh and eighth columns give the approximate $D$ or $\alpha$ values, estimated using $\bar{m}$-$\bar{s}$ plots.}
	\label{tab:cluster_results}
\end{table*}

\section{Discussion}
\label{sec:discussion}

\subsection{Comparison of methods}

We have shown that the BF star cluster model struggles to reproduce the observed features of substructured clusters. This because, as identified by JWL17, the BF model \emph{only} produces clusters with very high surface-density variances. Therefore, we strongly suggest that the model should be retired from star cluster analysis. The FBM model presented here overcomes this problem. FBM clusters have independent parameters which (i) control the amount of fractal clustering and (ii) set the global level of surface-density variation. In addition, clusters with $H\sim1$ fulfil the same role as centrally concentrated RDP clusters. This removes the need for using two unrelated models (i.e. BF and RDP) in the same analysis.

We find that $\mathcal{Q}$ and/or $\bar{m}$-$\bar{s}$ plot analyses are poorly suited to FBM clusters. Furthermore, they do not present robust uncertainties. This limits their efficacy, and we suggest that they should no longer be used as is. However, these analyses can be reformulated using modern machine learning techniques. Here, we present an ANN which makes robust estimates of star cluster parameters and their uncertainties. We note that alternatives to this method also exist. For example, JWL17 use principle component analysis to reduce a large range of observables to two principle features.

\subsection{Note on fractal dimension}

We have, where possible, avoided discussing these results in terms of fractal dimension. This is because $D$ does not uniquely describe a structure. For example, BF clusters with $D\sim3$ have a roughly uniform distribution of stars. Decreasing $D$ increases the level of substructure in the distribution. Conversely, FBM clusters with $D\sim3$ may be very substructured. Decreasing $D$ tends to fuse the sub-clumps together until the cluster is composed of one or two coherent objects. We therefore suggest caution when using the term ``fractal dimension'' in scientific statements. There appears to be little relation between the value of $D$ and the subjective \emph{clumpiness} expressed in different models.  

\subsection{Caveats}

While the FBM model has several advantages over the BF model, there are some caveats we must address. The FBM fields generated in Section \ref{sec:fbm_clusters} fill a periodic box with no true centre. Here, we shift the field's periodic centre of mass to the centre of the box. This generally places high density structures (if present) in the centre of the box, and low density regions around the edges. However, we acknowledge that this choice is arbitrary. Also, in some instances, the outline of the cluster can appear square (the effect is most pronounced when $\sigma$ is low; see Fig \ref{fig:clusters}). We could address this by culling the distribution into a sphere, but this would arbitrarily remove stars from the edges of the distribution.

We have demonstrated that the ANN regressor performs well at classifying and differentiating stars clusters. However, we note that there is an infinitude of measurable features for any given cluster. We have experimented with a large number of features from different graphs (e.g. centile-based statistics, features from the Delaunay triangulation). We have found through testing that the features presented here are adequate. Adding further features to the ANN only yields very minor improvements to its estimation accuracy.

\subsection{Future work}

The values of $H$ and $\sigma$ are useful for categorising star clusters by their morphology. However, in order to infer physical meaning from these measures, we need to apply them to simulations. We hypothesise that, for a sub-virial substructured cluster, $\sigma$ should increase as the cluster collapses under its own gravity. Meanwhile, $H$ should increase as the collapse erases the cluster's substructure. Conversely, for a super-viral equivalent of the same initial cluster, $\sigma$ should decrease over time as the cluster expands. It is not clear how $H$ will behave during this process. We will test these hypotheses by applying this analysis to an ensemble of $N$-body cluster simulations with various initial states.

The ANN method also provides a convenient way to compare the structure of the molecular clouds with that of star clusters. For example, \citet{ESS14} find that molecular clouds in the galactic plane typically have $H\lesssim0.4$, suggesting that they are similar in structure to Taurus, or Cha I. Previous numerical work has attempted to compare molecular cloud gas structure with that of embedded clusters \citep[e.g.][]{LWC11,PD15}. However, this was performed using $\mathcal{Q}$ analysis, which we have shown is unreliable. We will revisit this work and analyse the FBM properties of the stars and gas in molecular cloud simulations.

\section{Summary and conclusions}
\label{sec:conclusions}

We present an artificial star cluster model, based on Fractional Brownian Motion (FBM). The structure of these clusters is controlled by two parameters: the drift exponent $H$, which controls the degree of fractal structure, and the standard deviation $\sigma$ of the log-surface/volume density. The model is able to produce artificial clusters with a wide range of structural morphologies, similar to those of Lupus 3, IC 348, $\rho$ Oph, IC 2391, Cha I and Taurus. This contrasts with the Box-Fractal (BF) model -- used in $\mathcal{Q}$ analysis -- which has a single parameter, $D$. Here, $D$ is notionally a fractal dimension. However, changing its value simultaneously alters the degree of fractal structure and the amount of surface-density variation. Because these two properties are linked, the BF model is unable to reproduce naturally substructured clusters, like Cha I and Taurus. We note that \citet{JWL17} add extra parameters to the BF model in order to address this problem. Their model has a similar level of complexity as the FBM model, and can be viewed as an alternative to the work presented here.
 
$\mathcal{Q}$ analysis and $\bar{m}$-$\bar{s}$ plots are not well suited to estimating FBM cluster parameters. We present an Artificial Neural Network (ANN) regressor which can reliably estimate the parameter values and their uncertainties. Future work will involve using ANNs to measure how the structural properties of $N$-body cluster simulations evolve over time.  Further more, FBM analysis is well suited to studying the structure of the interstellar medium. This means we can use the method to directly compare the structure of gas and stars in star forming complexes.




\bibliographystyle{mnras}
\bibliography{refs} 

\section*{Acknowledgements}

\changes{We thank the reviewer, Simon Goodwin, for his constructive comments.} OL and APW gratefully acknowledge the support of a consolidated grant (ST/K00926/1) from the UK STFC. MLB gratefully acknowledges the support of a CDT in data intensive science (ST/P006779/1) from the UK STFC.

\appendix




\appendix

\section{Randomly sampling variates from 3 or 2 dimensional distributions}
\label{apn:random}

We can draw random coordinates $(X,Y,Z)$ from any gridded 3-dimensional distribution $p(x,y,z)$ using random variates $\mathcal{U}_x,\,\mathcal{U}_y\text{ and }\mathcal{U}_z$\changes{, drawn from the uniform distribution in the interval $[0,1]$}. First we calculate the cumulative distribution of $p(x,y,z)$ along the $x$-axis,
\begin{equation}
	\begin{split}
		P(x)&=\int\limits_{x_\textsc{min}}^x p(x)\,\mathrm{d}x\,,\\
		p(x)&=\int\limits_{z_\textsc{min}}^{z_\textsc{max}}\int\limits_{y_\textsc{min}}^{y_\textsc{max}}p(x,y,z)\,\mathrm{d}y\,\mathrm{d}z\,.
	\end{split}
\end{equation}
Here, the \textsc{min} and \textsc{max} subscripts denote the extreme coordinate values of the cartesian grid. Integrals are computed using the trapezium rule. Next, we numerically invert $P(x)$ to find $X$ using the relationship,
\begin{equation}
	\frac{P(X)}{P(x_\textsc{max})}=\mathcal{U}_x\,.
\end{equation}
In order to get $Y$, we calculate the cumulative distribution along the $y$-axis, given $X$,
\begin{equation}
	\begin{split}
		P(y|X)&=\int\limits_{y_\textsc{min}}^y p(y|X)\,\mathrm{d}y\,,\\
		p(y|X)&=\int\limits_{z_\textsc{min}}^{z_\textsc{max}}p(X,y,z)\,\mathrm{d}z\,.
	\end{split}
\end{equation} 
In practice, we precompute $P(y|x)$ for all gridded values of $x$. $P(y|X)$ is then found by linearly interpolating $P(y|x)$ over the two $x$-values either side of $X$. The $Y$-coordinate can found by inverting
\begin{equation}
	\frac{P(Y|X)}{P(y_\textsc{max}|X)}=\mathcal{U}_y\,.
\end{equation}
Finally, we get the $Z$ coordinate by calculating the cumulative distribution along the $z$-axis, given $X$ and $Y$,
\begin{equation}
	P(z|X,Y)=\int\limits_{z_\textsc{min}}^z p(z|X,Y)\,\mathrm{d}z\,,
\end{equation}
and inverting,
\begin{equation}
	\frac{P(Z|X,Y)}{P(z_\textsc{max}|X,Y)}=\mathcal{U}_z\,.
\end{equation}
Again, we precompute $P(z|x,y)$ for all combinations of $x$ and $y$, and bi-linearly interpolate $P(z|x,y)$ over the four $(x,y)$ values surrounding $(X,Y)$.  As before, the $Z$-coordinate is found by inverting
\begin{equation}
	\frac{P(Z|X,Y)}{P(z_\textsc{max}|X,Y)}=\mathcal{U}_z\,.
\end{equation}

This method can also be performed on a 2-dimensional distribution, $p(x,y)$. Here, we simply repeat the same steps (disregarding any integrals over the $z$-axis) until we have obtained $X$ and $Y$.

\section{Artificial neural network}
\label{sec:ANN}

An Artificial Neural Network (ANN) can be thought of as a collection of artificial neurons. Each neuron takes an input $\boldsymbol{x}=(x_1,x_2,\ldots,x_m)$ and outputs $z=f(b+\boldsymbol{w}\cdot\boldsymbol{x})$. Here, $b$ is a bias value, $\boldsymbol{w}$ is a vector of $m$ weights and $f(t)$ is an activation function. The activation function is usually chosen to vary smoothly over a limited range, e.g. $f(t)=\tanh(t)$ or $f(t)=1/[1+\exp(-t)]$. A collection of $n$ neurons can be grouped together to form a layer. Here, the weights are represented by an  $m\times n$ matrix $\boldsymbol{W}$, and the biases by a vector $\boldsymbol{b}$ with length $n$. The ensemble of neurons has an output $\boldsymbol{z}=f(\boldsymbol{b}+\boldsymbol{W}\boldsymbol{x})$.\footnotemark

\footnotetext{Note that here $f(t)$ is a scalar function with a scalar argument. For the same function, we define $f(\boldsymbol{t})\equiv(f(t_1),f(t_2),\ldots)$.}

For this analysis, we set up a three-layer ANN using the \textsc{MLPRegressor} class in the \textsc{Scikit-Learn} library \citep{scikit-learn}.\footnote{The hyperparameters of the class, including the number of layers and neurons per layer, are tuned using \textsc{GridSearchCV} cross-validation tool. The full implementation can be found at \texttt{github.com/odlomax/clusterfrac}.} The structure of the ANN is as follows:
\begin{equation}
	\begin{array}{ll}
		\text{Layer 1, $m$ elements:} & \boldsymbol{x}\,; \\
		\text{Layer 2, $n$ elements:} & \boldsymbol{z}=\tanh(\boldsymbol{b}_1+\boldsymbol{W}_1\,\boldsymbol{x})\,; \\
		\text{Layer 3, $p$ elements:} & \boldsymbol{y}=\boldsymbol{b}_2+\boldsymbol{W}_2\,\boldsymbol{z}\,.
	\end{array}
	\label{eqn:ann}
\end{equation}
The first layer is the input vector of features $\boldsymbol{x}$. This is composed of the measurable properties of a star cluster. The second layer $\boldsymbol{z}$ is determined by a bias vector $\boldsymbol{b}_1$ and the weight matrix $\boldsymbol{W}_1$. The third and final layer is the output $\boldsymbol{y}$. This is composed of the underlying cluster parameters which we are trying to estimate. The values are determined by a second bias vector $\boldsymbol{b}_2$ and weight matrix $\boldsymbol{W}_2$. Note that no activation function is used to calculate the final layer; this is so $\boldsymbol{y}$ is not confined to a limited range.  The number of neurons in the second layer is arbitrary; here, we find $n=40$ provides \changes{the most accurate results} (more complicated ANN regressors may contain multiple hidden layers). For simplicity, we refer to the ANN mapping of $\boldsymbol{x}$ to $\boldsymbol{y}$ as $\boldsymbol{y}=F(\boldsymbol{x})$. The ANN is trained by taking $N_\textsc{train}$ training clusters, with known $\boldsymbol{y}$, and finding values of $\boldsymbol{W}_1$, $\boldsymbol{W}_2$, $\boldsymbol{b}_1$ and $\boldsymbol{b}_2$ which minimise $\langle (F(\boldsymbol{x})-\boldsymbol{y})^2\rangle$. This is performed by the class using gradient-descent techniques.


\bsp	
\label{lastpage}
\end{document}